\documentclass[aps,amsmath,notitlepage,twocolumn,amssymb,prb,longbibliography]{revtex4-1}

\usepackage{hyperref}
\usepackage[utf8]{inputenc}
\usepackage{bm}
\usepackage{verbatim}
\usepackage{subfigure}
\usepackage{graphicx}
\usepackage{amsmath}
\usepackage{color}
\usepackage{float}
\usepackage{bm}
\usepackage{amssymb}
\usepackage{slashed}
\usepackage{wasysym}
\begin{document}

\title{Infinite critical boson induced non-Fermi liquid in $d=3-\epsilon$ dimensions}
\date{\today}
\author{Zhiming Pan}
\affiliation{Institute of Natural Sciences, Westlake Institute for Advanced Study, Hangzhou, China}
\affiliation{Department of Physics, School of Science, Westlake University, Hangzhou 310024, China}
\author{Xiao-Tian Zhang}
\email{zhangxt@hku.hk}
\affiliation{Department of Physics, the University of Hong Kong, Hong Kong SAR, China}
\date{\today}

\begin{abstract}
We study the fermion-boson coupled system in $d=3-\epsilon$ space dimensions near the quantum phase transition;
infinite many boson modes locating on a sphere become critical simultaneously, which is dubbed ``critical boson sphere'' (CBS).
The fermions on the Fermi surface can be scattered to nearby points locating on a boson ring in the low-energy limit.
The number of boson scattering channel $N_{B}$ is also infinite,
which renders the well-known Landau damping effect largely suppressed.
The one-loop renormalization group analysis is performed with asymptotic $\epsilon$-expansion.
We find that the fermion self-energy and Yukawa interaction vertex are dressed with $\epsilon$ poles;
in addition, there emerges an enhancement due to the curvature effect of CBS.
In certain perturbative regime, we identify a marginal non-Fermi liquid (NFL) fixed point
that exists intrinsically in the large-$N_B$ limit.
The infinite critical bosons comprise a physical realization of the flavor degrees of freedom
which has been proposed for matrix large-$N_B$ bosons.
\end{abstract}

\maketitle

\section{Introduction}
\label{sec1}

Quantum phase transitions in strongly correlated metals
comprise a challenging and intriguing area of studies in modern condensed matter physics.
Multiple physical degrees of freedom emerge in the vicinity of the quantum critical point:
the gapless fermionic quasi-particles and critical fluctuation of bosonic order parameters.
The coupling between these gapless matters render the Landau Fermi liquid picture invalid intuitively,
yet it presents serious subtlety for theorists;
various theoretical methods are developed to address the issue of non-Fermi liquid (NFL) behavior,
including the perturbative scheme in $\epsilon$-expansion and/or the large-$N$ theories\cite{PhysRevLett.74.1423,Nayak1994,Nayak1994B,Senthil2009,Metlitski2010,Mross2010,Gonzalo2012,SSLee2013,Raghu2013,Raghu2013B,Gonzalo2014,Gonzalo2015,shi2022gifts}.

A particular type of the transitions, known as continuous Pomeranchuk transition,
is extensively studied where order parameter condenses
at a zero momentum point and coupled to each point on the Fermi surface.
Recently, there has been heated investigations on the critical boson field (or order parameters)
with continuously degenerate dispersion minima.
The stable existence of a critical boson surface has been studied in the context of 
Bose Luttinger liquids\cite{Yang2019,Lake2021} and the so-called ``Bose metal''\cite{Paramekanti2002,Phillips2003,Motrunich2007}.
Various NFL-behaviors are proposed by coupling fermions to the critical boson surfaces\cite{Lake2021,YBKim2021,Zhang2021,Tsvelik2021,Ku2021}.
Conventional theoretical treatments focus on the deep infrared regime around the quantum critical point,
which then provides information for the mediated finite temperature sector. 
To get access to the quantum criticality, a well-established method proposed by 
\emph{J. Hertz}\cite{Hertz1976} and \emph{A. J. Millis}\cite{Millis1993} relies on integrating out the gapless fermions first;
as a result, a singular polarization function is generated for bosons 
which alters the spacetimes scaling.
In a previous paper\cite{Zhang2021}, one of us follows this line of thinking and 
develops a Hertz-Millis theory for Fermi surface coupled to 
the critical boson surface (CBS) in $d=3+1$ dimensions.

More straightly, the finite temperature NFLs can be studied starting 
from a weakly coupled metals at intermediate scale.
A systematic Wilsonian renormalization group (RG) theory has been developed
where the fermions and bosons are treated on equal-footing\cite{Raghu2013B,Gonzalo2014,Gonzalo2015};
instead of integrating over fermions at all energy scale,
only high energy fermion excitations are integrated out.
The remaining system comprise both fermion and boson degree of freedom.
In this paper, we are motivated to study fermions coupled to boson with
critical boson surface at finite energy regime. 
In the presence of CBS, a given fermion on the fermion surface
can be scattered by infinite many critical bosons on the CBS,
which amounts to a large-$N_B$ limit\cite{Raghu2013,Raghu2014}.
We develop a controlled perturbative theory in asymptotic $\epsilon$-expansion 
around the critical spatial dimensions $d=3-\epsilon$.
An one-loop RG analysis of fermionic self-energy and vertex is carried out by evaluating the quantum corrections coming from the Yukawa coupling between fermions and boson.
We discover a novel type of NFL fixed point owing to the existence of infinite critical bosons.

The outline of the paper is as follow.
We first introduce a fermion-boson coupled system with a finite Fermi surface and critical boson surface in Sec.~\ref{sec2}.
Focusing on the low-energy behavior, we analysis the relevant degree of freedom when fermion on the Fermi surface is scattering through critical boson near CBS.
The effective theory can be further characterized by an intrinsic large-$N_B$ formulation where the Landau damping effect is highly suppressed. 
Following the effective large-$N_B$, we study the scaling analysis of the fermion and boson.
Motivated by the marginal dimension $d=3$ of the Yukawa coupling, 
we set up a renormalization group procedure within $d=3-\epsilon$ dimensional regularization scheme.
In Sec.~\ref{sec3}, we perform an one-loop RG analysis for the fermion self-energy and Yukawa interaction.
We derive the one-loop $\beta$-function of the coupling parameter and the anomalous dimension of fermion.
We discover a NFL fixed point solution existing particularly in the large-$N_B$ limit 
and classify its property by comparing to the usual $N_B=1$ counterpart.
In the final Sec.~\ref{sec4}, we summarize the results and clarify the controlled energy regime
where the present effective theory fits. And, we comment on other energy regimes
and potential directions for future works.

%--------------------------Fig-------------------------------
\begin{figure}[htbp] 
	\centering
	\includegraphics[width=6.5cm]{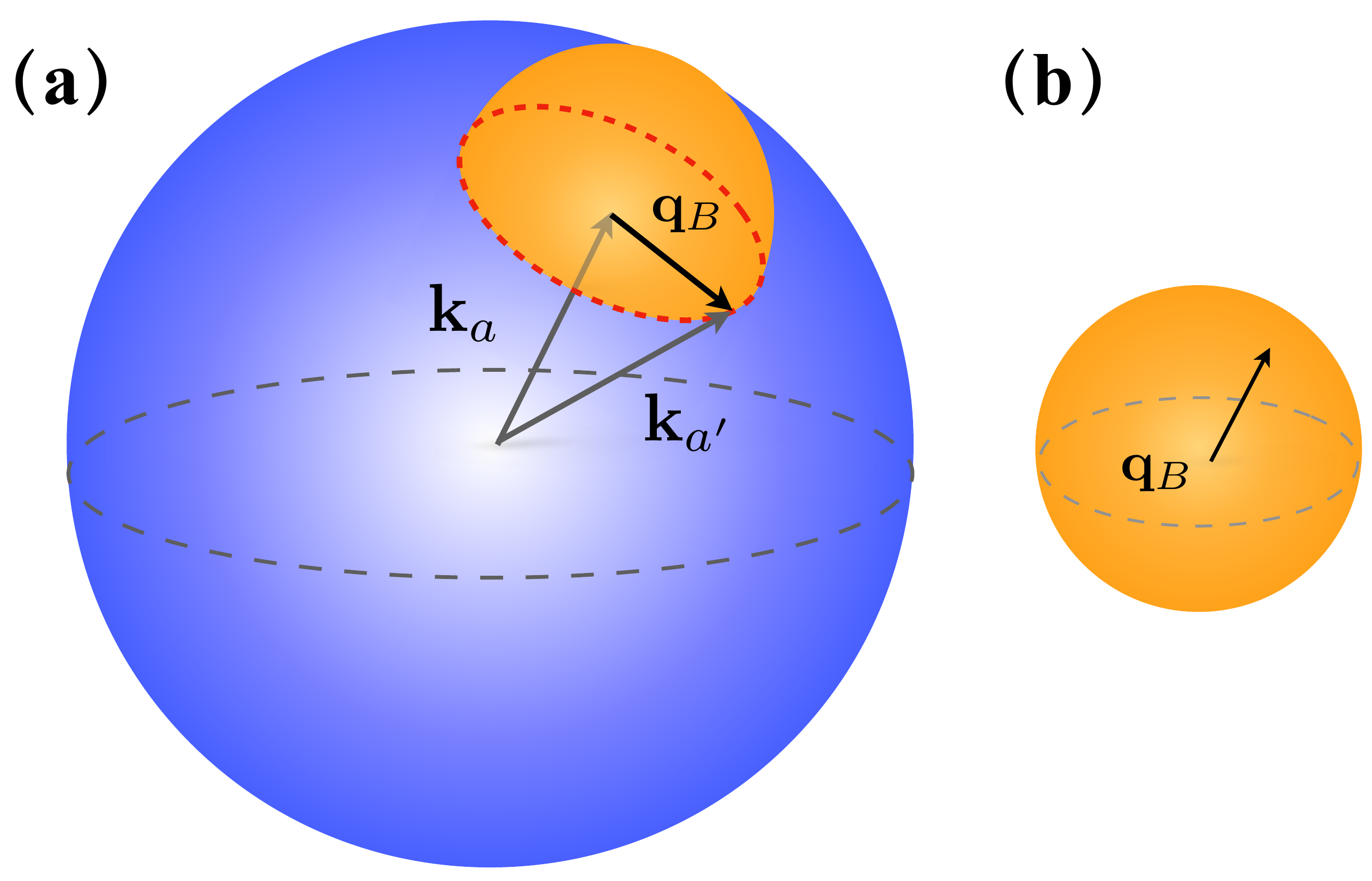}
	\caption{(Color online.) (a) The blue sphere is the 
	Fermi surface for the Fermi liquids.
	The orange hemisphere represents
	the critical boson surface.
	${\bf k}_a=k_F\hat{k}_a,{\bf k}_{a^{\prime}}$ is the Fermi momentum, 
	and $q_B$ is the radius of the boson surface.
	The Yukawa coupling connects fermions 
	with a ring	of fermion modes on the Fermi surface.
	(b) The critical boson sphere with boson momentum ${\bf q}_B =q_B\hat{q}_B$. 
	}
	\label{Fig1}
\end{figure}
%------------------------------------------------------------

\section{Effective action for the fermion-boson coupled system}
\label{sec2}

We consider an action in $d+1$ Euclidean spacetimes consisting 
respective low-energy effective theories for fermions and bosons, 
and with a Yukawa interaction between them:
\begin{equation}
\begin{aligned}
{\cal S} = & {\cal S}_{f}+ {\cal S}_{b} + {\cal S}_{bf}, \\
{\cal S}_{f} = & \int \frac{d^{d+1}k}{(2\pi)^{d+1}} ( ik_0 - \xi_{\bf k} )\psi^\dagger(k)\psi(k),\\
{\cal S}_{b} = & \int \frac{d^{d+1}q}{(2\pi)^{d+1}} \big[ r + q_0^2 + (|{\bf q}|-q_B)^2 \big] \phi(-q)\phi(q),\\
&  + u \int d^{d+1}x \ \phi^4(x), \\
{\cal S}_{fb} = &\int \frac{d^{d+1}k}{(2\pi)^{d+1}}\int \frac{d^{d+1}q}{(2\pi)^{d+1}} 
g(k,q)\phi(q) \psi^\dagger(k) \psi(k+q), \\
\end{aligned}
\label{eq1}
\end{equation}
where $k=(k_0,{\bf k})$ and $q=(q_0,{\bf q})$ are $d+1$ dimensional momenta.
The first term ${\cal S}_{f}$ describes a standard Landau Fermi liquid of spinless $\psi$ field,
with a quadratic band dispersion $\xi_{\bf k}= {\bf k}^2/2m - \mu$ and Fermi level $\mu$.
The second term ${\cal S}_{b}$ represents collective boson modes $\phi$
whose spectrum ${\cal E}^2({\bf q})=(|{\bf q}|-q_B)^2$ have a continuous degenerated minima on a sphere $|{\bf q}|=q_B$.
$u$ is the strength of quartic boson self-interaction.
Tuning the parameter $r$, the boson modes on the sphere
reach criticality simultaneously, which is termed as ``critical boson surface'' (CBS).
The third term is the Yukawa coupling between
the scalar boson and fermion fields in the momentum space. 
The Yukawa coupling is in general a function form $g(k,q)$, measuring the scattering process while a fermion with momentum ${\bf k}$ is scattered into ${\bf k+q}$ through the boson mode ${\bf q}$.

\subsection{Low-energy configuration for boson-fermion coupled system}
\label{sec2a}

\begin{figure}[htbp] 
	\centering
	\includegraphics[width=7cm]{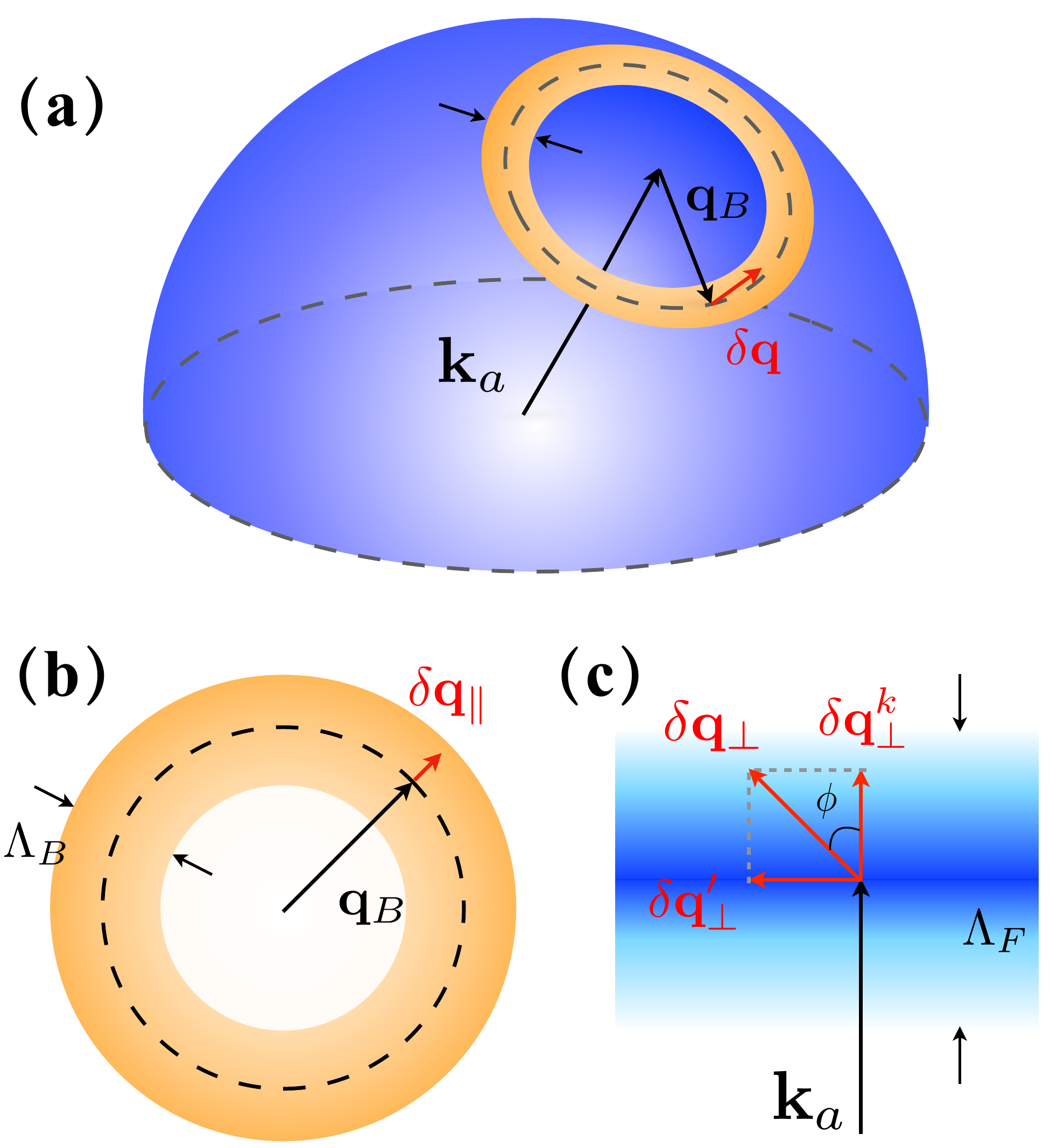}
	\caption{(Color online.) 
	Schematic picture for the momentums decomposition in the low-energy scattering process.
	(a) CBS located at fermi momentum ${\bf k}_a$ intersects with the Fermi surface at the ring shell region plotted in orange.
	(b) and (c) are the vertical and cutting views of the momentum configuration respectively.
	(b) Momentum shell of the critical boson ring along parallel direction.
	The parallel component $\delta {\bf q}_\parallel$ is bounded by $\Lambda_B$;
	(c) The perpendicular momentum components projected to the ${\hat q}_B$-plane.
	$\delta {\bf q}$ is decomposed with respect to the $\hat{k}_a$-axis,
	the components along/perpendicular to the the axis 
	is denoted as $\delta {\bf q}_\perp^k/\delta {\bf q}_\perp^\prime$.
	$\delta {\bf q}_\perp^k$ represents the displacement off the Fermi surface (dark blue color)
	which is illustrated by the gradually fading blue colors. The momentum shell is bounded by $\Lambda_F$
	and the 2D polar angle in the perpendicular plane is denoted as $\phi$.
	}
	\label{Fig2}
\end{figure}

The low-energy effective theory of the coupled system near the quantum criticality 
is dominated by the physics near the Fermi surface and CBS.
To be precise, the low-energy fermion ${\bf k}$ is restricted to a momentum shell $\Lambda_F$ near Fermi surface, 
$k_F-\Lambda_F<|{\bf k}|<k_F+\Lambda_F$.
The fermion can be linearized around Fermi surface ${\bf k}_F=k_F\hat{k}_F$ with respect to $\delta {\bf k}= {\bf k}-{\bf k}_F$.
The low-energy fermion dispersion reads $\xi_{\delta {\bf k}}\simeq v_F \hat{k}_F \cdot \delta {\bf k}$ with  fermion velocity $v_F =k_F/m$.
In another word, the momentum near Fermi surface is uniquely expressed by the solid angle $\hat{k}_F$
and a radical displacement $\hat{k}_F\cdot \delta{\bf k}$,
which is also known as the spherical scaling scheme\cite{Shankar1994}.
Similarly, the boson momentum can be expanded 
around the vicinity of CBS ${\bf q}_B=q_B\hat{q}_B$ by defining 
a small variation $\delta {\bf q} ={\bf q}- {\bf q}_B$ as illustrated in Fig.~\ref{Fig2}(a).
It is reasonable to assume that the radius of the boson sphere 
is much smaller than fermion one, namely we have $q_B\ll k_F$.
A given fermion momentum ${\bf k}$ on the Fermi surface, 
mediated by the boson mode with a momentum ${\bf q}$, 
is transferred to nearby points ${\bf k+q}$ near the Fermi surface.
The CBS is attached to the Fermi surface as demonstrated in Fig.~\ref{Fig1},
therefore, the curvature effect of the CBS is more prominent for the low-energy physics.
In contrast to the linearization around Fermi surface,
we retain the quadratic terms around the CBS
which is a manifestation of the curved boson sphere in vicinity of ${\bf q}_B$.
We note that the curvature effect brought by the quadratic term
is crucial for the one loop RG as dictated in Sec.~\ref{sec3}.
The boson dispersion is approximated as
\begin{equation}
\begin{aligned}
|{\bf q}|-q_B \simeq {\cal E}(\delta {\bf q}) \equiv \delta q_\parallel + \frac{\delta q_\perp^2}{2q_B} \ ,
\end{aligned}
\label{eq5}
\end{equation}
where we have decompose the small variation around CBS
into parallel and perpendicular components.
The parallel direction is aligned along the radical $\hat{q}_B$-direction
and the perpendicular components lies in the tangential plane,
which are illustrated in Fig.~\ref{Fig2}(b) and (c) respectively.
Namely, we have
\begin{equation}
\delta {\bf q}_\parallel= (\hat{q}_B \cdot \delta {\bf q})\hat{q}_B, \ \ 
\delta {\bf q}_\perp = \delta {\bf q}- \delta {\bf q}_\parallel.
\label{eq2}
\end{equation}
The parallel and perpendicular components 
are bounded by two independent UV cutoffs 
$|\delta q_\parallel|<\Lambda_B, |\delta q_\perp|<\Lambda_F$.
Here $\Lambda_F/\Lambda_B$ regulates the scaling towards 
the nearest point on the Fermi surface/CBS respectively.

 %--------------------------Fig-------------------------------
%------------------------------------------------------------

The low-energy dynamics of the fermion-boson coupled system
is described by the inelastic scattering process around Fermi surface.
The Yukawa coupling function admits a Taylor expansion
\begin{equation}
g(k,q) \simeq g({\bf k}_F,{\bf q}_B) + g_1 \delta k + g_2 \delta q,
\end{equation}
with small momentum variations $\delta k\ll k_F, \delta q\ll q_B$.
The leading order term $g({\bf k}_F,{\bf q}_B)$ describes 
an inelastic scattering process for fermions on the Fermi surface,
which is depicted in Fig.~\ref{Fig1}(a).
The momentum transfer is facilitated by the bosons on the CBS
which intersects with Fermi surface in a ring shaped region, which is dubbed ``critical boson ring''.
Without energy transfer, a given Fermi momentum point ${\bf k}={\bf k}_F$ can be scattered 
back to the ring on the Fermi surface.
Owing to the small boson sphere $q_B/k_F\ll 1$,
the critical boson ring is approximately perpendicular to the Fermi momentum ${\bf k}_F$.
This configuration motivates a further decomposition for the perpendicular component $\delta {\bf q}_\perp$,
which reads [see Fig.~\ref{Fig2}(c)]
\begin{equation}
\delta {\bf q}_\perp^k = ({\hat k}_F\cdot \delta{\bf q}){\hat k}_F, \ \ \delta {\bf q}_\perp^\prime = \delta {\bf q}_\perp - \delta {\bf q}_\perp^k.
\label{eq3}
\end{equation}
Eq.(\ref{eq3}) together with Eq.(\ref{eq2})
establish a relation between boson and fermion momentum variations,
namely the fermion momentum can be viewed as a perpendicular component for the boson's.
Owing to the presence of CBS, the entire coupled system 
has to be incorporated into a consistent set of scaling rules.
This momentum configuration comprises an essential feature for the fermions coupled to CBS.
In many previously studied boson-fermion coupled system\cite{Raghu2013,Gonzalo2015},
boson modes are on the verge of condensation only at a countable number of momentum points
while the fermions live on the Fermi surface.
The bosons and fermions have completely different kinematics,
thus, different scaling schemes are adopted.
For instance, the fermions can be described by the conventional spherical scaling 
whereas all components of boson momentums have to be included\cite{Gonzalo2014}.
The low-energy fermions can exchange high energy bosons,
which could result into a ``UV-IR mixing'' and introduces divergence for the quantum corrections
in the RG analysis\cite{Gonzalo2015}.

\subsection{Effective large-$N_B$ theory}

The critical boson ring provides multiple scattering channels for fermions in the low energy,
which amounts to a large boson flavors $N_B$ with the fermion flavor kept at unity $N_F=1$.
This situation is in sharp contrast to the conventional anti-ferromagnetic transition,
where the ordering is at a single finite wave-vector (or a countable number of wave-vectors at most).
In the large-$N_B$ limit, the bosons act as the dissipative bath for the fermions,
thus the fermionic quasi-particles are strongly affected 
which usually leads to the breakdown of the Landau Fermi liquid picture.

The large $N_B$ is an interesting alternative description to the usual large-$N_F$ theories \cite{Gonzalo2014,Yang2019,Damia2019,shi2022gifts}.
In the conventional large-$N_F$ theories near quantum transitions, 
the number of fermion flavors is much larger than the boson's.
The fermions and bosons are not treated on equal-footing,
instead, the priority of the coupled system is to evaluate 
the boson polarization induced by the large number of fermions.
The dominant feature is the damping effect of the bosons brought 
by the fermion particle-hole excitation around Fermi surface.
Historically, quantum metals in the presence of finite number of critical bosons
are well-described by the Hertz model (also known as the Hertz-Millis-Moriya method)\cite{Hertz1976,Millis1993}; 
integrating out the gapless fermions gives rise to $\sim g^2 |q_0|/|{\bf q}|$.
The Hertz term is singularly dependent on the momentum,
which is the price paid during integrating out gapless matters
and is not compatible with Wilsonian RG.
This peculiar momentum and frequency dependence, if relevant,
alters scaling dimensions of the spacetimes for the original low-energy theory in Eq.(\ref{eq1}).
\emph{S. Raghu} et al. developed a Wilsonian effective field theory\cite{Raghu2013}, 
where the Hertz term is not included from the beginning;
rather, it is generated as one-loop correction at ${\cal O}(g^2)$.
The leading contribution is still provided by the bare boson terms 
above a characteristic energy scale $\omega_{Ld}\sim {\cal O}(g)$.
Approaching the lower energies $\omega < \omega_{Ld}$,
the Landau damping effect gradually sets in.
In such a critical regime, Hertz method has been used to 
address the non-Fermi liquid behavior in the presence of the CBS\cite{Zhang2021}.
In the present study, we focus on the finite energy regime $\omega>\omega_{Ld}$;
our goal is to establish non-singular effective theory
and explore possible perturbative NFL fixed points.

Up to date, only a few works\cite{Raghu2013,Raghu2013B,Raghu2014,Gonzalo2014} 
have been devoted for the opposite large-$N_B$ limit.
The fermion/boson flavors are generalized to larger values
by allowing a larger fundamental/adjoint representation.
The large-$N_B$ is introduced by hand to suppress the characteristic energy scale $\omega_{Ld}$
and to avoid various subtleties associated with four-Fermi interactions.
This flavor degree of freedom is expected to be relevant for strongly correlated electronic systems,
particularly, the high-$T_c$ superconducting materials.
In our case, the large-$N_B$ degree of freedom shows up in a natural way:
infinite many critical bosons on the critical boson ring
provides multiple scattering channels for fermions,
thereby, serves as a physical realization for large-$N_B$ quantum critical metals.

 %--------------------------Fig-------------------------------
\begin{figure}[htbp] 
	\centering
	\includegraphics[width=6cm]{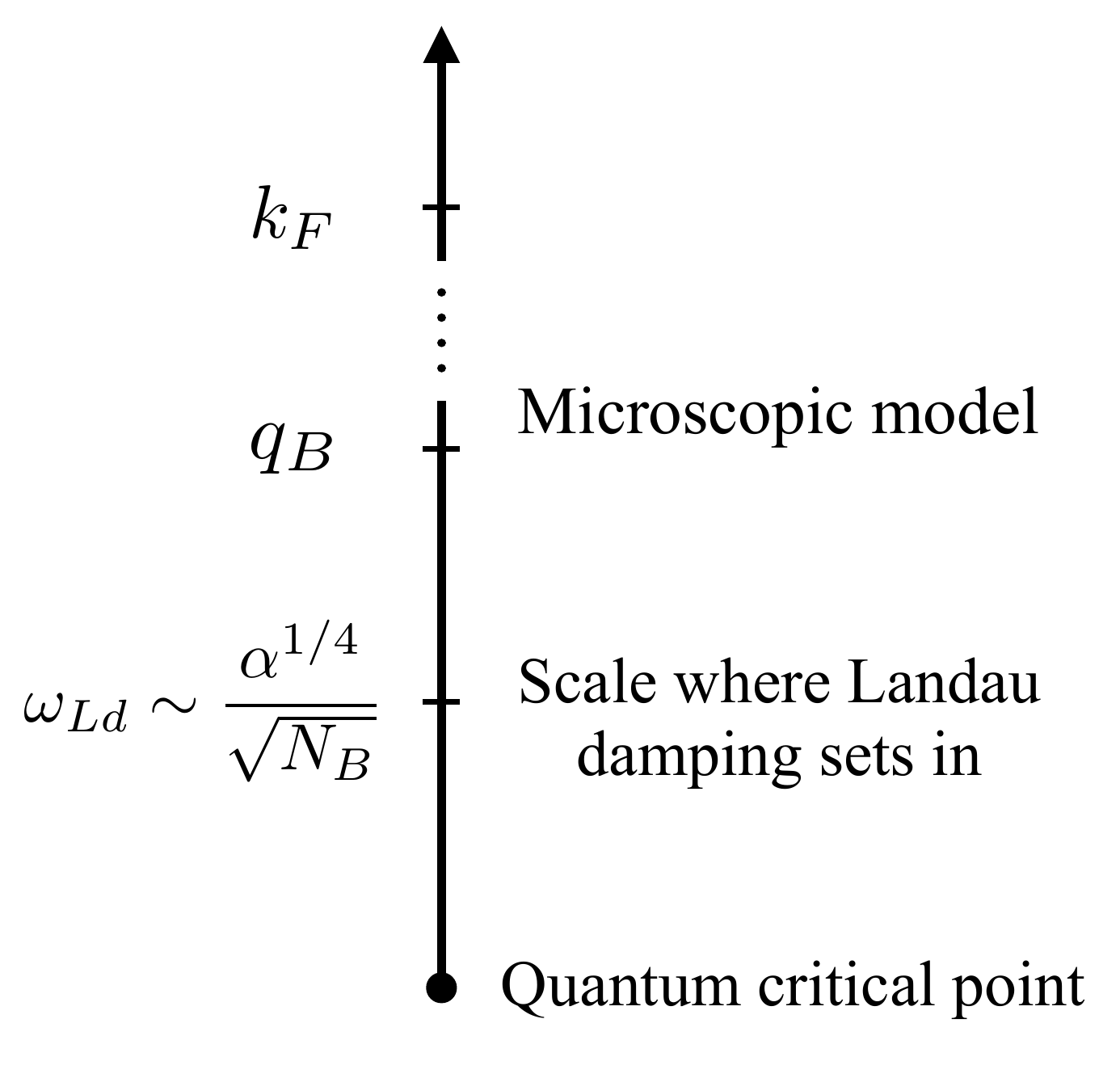}
	\caption{
	The relevant energy scales for the Fermi surface and CBS coupled system.
	The microscopic energy scales are set by the radius of CBS $q_B$ and Fermi sphere $k_F$.
	The controlled regime for the low-energy effective theory applies below the microscopic energies
	and above the Landau damping characteristic energy $\omega_{Ld}$.
	Below $\omega_{Ld}$, the Landau damping effect gradually sets in
	and the Hertz-Millis theory is considered to be a legitimate description.
	}
	\label{Fig3}
\end{figure}
%------------------------------------------------------------

To be specific, we write the fermions and bosons 
to be vector and matrix fields respectively $\psi_a, \phi_{aa^\prime}$.
Here $a$ is the index for the momentum points ${\bf k}_a=k_F\hat{k}_a$ on the Fermi surface, 
and $a^\prime$ represents points ${\bf k}_{a^{\prime}}=k_F\hat{k}_{a^{\prime}}$ locating on the
the critical boson ring centered at $a$ as plotted in Fig.~\ref{Fig1};
the difference between $a$ and $a^{\prime}$ is the boson momentum ${\bf q}_{B}={\bf k}_{a^{\prime}}-{\bf k}_a$.
The fermion-boson coupled system is rotational symmetric, which leads to a coupling constant $g_0\equiv g(k_F,q_B)$ depending only on the absolute value of Fermi and boson momentum.
We further omit the dependence on the large boson momentum ${\bf q}_B$ for simplicity and 
proceed with a low-energy effective theory for the quantum critical metal at $r=0$, which reads
\begin{equation}
\begin{aligned}
{\cal S}_{f} =& \int \frac{d^{d+1}\delta k}{(2\pi)^{d+1}}(ik_0 - v_F \delta k_\perp^k )\psi^\dagger_a(\delta k)\psi_a(\delta k),\\
{\cal S}_{b} =& \sum_{a^\prime \in O_a}\int \frac{d^{d+1}\delta q}{(2\pi)^{d+1}}\big[q_0^2 + {\cal E}^2(\delta {\bf q}) \big] 
 \phi_{a^\prime a}(-\delta q)\phi_{aa^\prime}(\delta q),\\
{\cal S}_{fb} =& g_0 \sum_{a^\prime \in O_a} \int \frac{d^{d+1}\delta k}{(2\pi)^{d+1}}\int \frac{d^{d+1}\delta q}{(2\pi)^{d+1}} 	\\
&\qquad\qquad \times\phi_{aa^\prime}(\delta q) \psi^\dagger_a(\delta k) \psi_{a^\prime}(\delta k+\delta q). \\
\end{aligned}
\label{eq11}
\end{equation}
Here $O_a$ labels the critical boson ring centered at $a$ and
the summation $a^\prime$ is over the $N_B$ number of scattering channels.
$\delta q\equiv (\delta {\bf q},q_0)$ is the small variation for the four-momentum ${\bf q}$ of boson while $\delta k\equiv (\delta {\bf k},q_0)$ for the fermion.

\section{Renormalization scheme}

Based on the low-energy effective theory,
we first develop a consistent scaling procedure for the momentums of bosons and fermions.
Then, we study the perturbative scheme for $\epsilon$-expansion in $d=3-\epsilon$ dimensions.

\subsection{Tree level scaling analysis}

Starting from the effective action in Eq.(\ref{eq11}),
we adopt the following scaling scheme:
\begin{equation}
\begin{aligned}
q_0^\prime = e^{zl }q_0, \ \ \delta {\bf q}_\parallel^\prime = e^{l} \delta {\bf q}_\parallel, \ \ 
\delta {\bf q}_\perp^\prime = e^{\gamma l} \delta {\bf q}_\perp, \\
\end{aligned}
\label{eq4}
\end{equation}
where the parallel momentum is chosen to be the primary component,
and the perpendicular one has an extra factor $\gamma \le 1$ being determined.
The dynamical exponent remains to be $z=1$,
since we have ignored the Landau damping term.
The scaling relations in Eq.(\ref{eq4}) leads to,
\begin{equation}
\begin{aligned}
& {\rm dim}[\phi^2(\delta q)] = -[3+z+(d-1)\gamma], \\ 
& {\rm dim}[\psi^2(\delta k)] = -[z+1+(d-2)\gamma] .\\
\end{aligned}
\end{equation}
As a result, the four-fermion coupling constant has a scaling dimension $-(z+1+d) + 3\gamma<0$,
which is irrelevant and consistent with the Fermi liquid theory.
The boson quartic interaction and Yukawa coupling are endowed with scaling dimensions,
\begin{equation}
\begin{aligned}
& {\rm dim}[u] = 3-z- (d-1)\gamma \\
& {\rm dim}[g_0] = -\big[ (z-1) + (d-3)\gamma\big]/2 \\
\end{aligned}
\end{equation}
The Yukawa coupling constant is marginal when $z=1,d=3$
regardless of the value of $\gamma$,
which is not the case for the boson quartic interaction.

We are in the position to fix the value of $\gamma$,
which is mainly determined by how we deal with the curvature term in Eq.(\ref{eq5}).
We prefer to have the same scaling dimensions for all components of momentum
due to the special momentum configuration illustrated in Sec.~\ref{sec2a};
yet, this would render the curvature term irrelevant.
We pursue a scaling scheme
where the curvature effect of the CBS can be included
while keeping all components of momentums to be unity.
Recall that the parallel and perpendicular components
have separate UV cutoffs {\sl i.e.} $|\delta q_\parallel|<\Lambda_B, |\delta q_\perp|<\Lambda_F$,
we can make a further approximation for the curvature term in the boson dispersion
\begin{equation}
\begin{aligned}
{\cal E}^2(\delta {\bf q}) \simeq \delta q^2_\parallel + \alpha \delta q_\perp^2, \ \ \ \alpha =\Lambda_B/q_B \ll 1.
\end{aligned}
\label{eq6}
\end{equation}
The small constant $\alpha$ parametrizes the curvature of the boson sphere.
we will see in following section that the approximation captures the main feature in one-loop RG
and greatly simplifies the calculations in a reasonable manner.
Eq.(\ref{eq6}) suggests that $\gamma=1$
and the boson quartic interaction and Yukawa coupling are both marginal at classical dimension $d=3$.
We are motivated to find a perturbative fixed point
in an $\epsilon$-expansion with $d=3-\epsilon$.

\subsection{Renormalization procedure in the $\epsilon$-expansion}
The appearance of the Fermi surface highly modifies the renormalization of the theory.
We study the quantum theory in asymptotic $\epsilon$-expansion of field renormalization,
which differs from the conventional Wilsonian RG scheme.
We follow the renormalization procedure adopted in literature\cite{Gonzalo2014}
and briefly review it in the following. 

The bare quantities (denoted with subindex $0$) in the original theory 
can be related to the physical (renormalized) quantities through the renormalization factors,
\begin{equation}
\begin{aligned}
& \psi_0 = Z_\psi^{1/2} \psi, \ \ \phi_0 = Z_\phi^{1/2} \phi, \\
& g_0 = \mu^{\epsilon/2}\frac{Z_g}{Z_\psi Z_\phi^{1/2}} g, \ \ v_F^0 = Z_v v_F, \\
\end{aligned}
\label{eq10}
\end{equation}
where $\mu$ is an arbitrary RG energy scale and $g$ is the dimensionless coupling constant.
The Lagrangian density of the original theory becomes
\begin{equation}
\begin{aligned}
{\cal L} = & Z_\phi \phi(q_0^2+{\cal E}^2)\phi + Z_\psi\psi^\dagger\big(ik_0- Z_v v_F\delta k_\perp^k \big)\psi  \\
& + \mu^{\epsilon/2} Z_g g \phi \psi^\dagger \psi,\\
\end{aligned}
\label{eq13}
\end{equation}
where we have abbreviated the function variables. Now, we are in the position
to separate the renormalization factors into the counter-terms
\begin{equation}
\begin{aligned}
&Z_\psi = 1 +\delta_\psi, \ \ Z_\phi =1 +\delta_{\phi}, \\ 
&Z_\psi Z_v= 1+\delta_v, \ \ Z_g = 1+\delta_g.
\end{aligned}
\end{equation}
To the one-loop level, the boson polaization diagram Fig.~\ref{Fig4}(b) is finite and $\delta_{\phi}=0$.
Using these relations, and differentiating both sides of Eq.(\ref{eq10})
with respect to the RG scale $\mu$, we obtain the beta functions
and fermion anomalous dimension,
\begin{equation}
\begin{aligned}
& \gamma_{\psi}= \frac{1}{2}\frac{d\delta_\psi}{d\ln\mu}, \\
& \beta_v = 2\beta v_F-\frac{d\delta_v}{d\ln\mu},\\
& \beta_g = \Big(-\frac{\epsilon}{2} + 2\gamma_{\psi}-\frac{d\delta_g}{d\ln\mu}\Big)g. \\
\end{aligned}
\label{eq14}
\end{equation}

The counter-terms defined above are physical important
in the sense that they cancel the divergent $\epsilon$-poles from the fermion self-energy $\Sigma(k)$
and the Yukawa interaction vertex $\Gamma$ at one-loop renormalization. Explicitly, we have the following equations for the fermion propagator $G$ and Yukawa vertex,
\begin{equation}
\begin{aligned}
& -\big[G(k)-G_0(k)\big] = ik_0 \delta_\psi - \delta_v v_F \delta k_\perp^k + \Sigma(k) \\
& {\cal L}_{fb} = \mu^{\epsilon/2}g\big[1+\delta_g+\Gamma(k;q)/g\big] \phi(q)\psi^\dagger(k)\psi(k+q). \\
\end{aligned}
\end{equation}
We note that $\Gamma(k;q)$ here is generally momentum and frequency dependent;
moreover, the dependence is usually non-local in the quantum critical metal
[namely the operator diverges as the momentum/frequency approaches low-energy, long-wavelength limit]\cite{Gonzalo2014,Gonzalo2015}.
A conservative way would be extracting the constant part of the $\Gamma$ function
which can be cancelled by a simple counter-term $\delta_g$.
We adopt a more careful treatment for the non-local vertex renormalization in Sec.~\ref{sec3}.

\section{Perturbative analysis in $d=3-\epsilon$ dimensions}
\label{sec3}

\begin{figure}[htbp] 
	\centering
	\includegraphics[width=7cm]{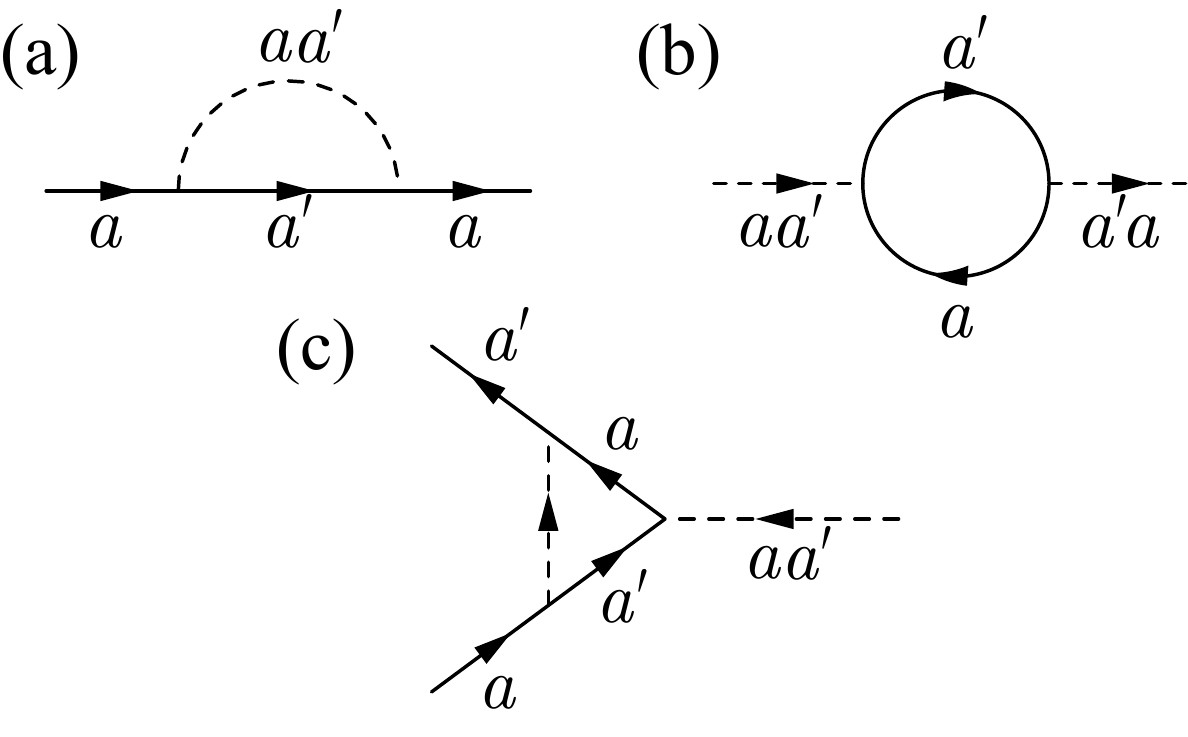}
	\caption{ One-loop Feynman diagrams of (a) fermion self-energy, 
	(b) boson polarization and (c) Yukawa vertex. 
	The dashed/solid line denotes the boson/fermion propagator.
	}
	\label{Fig4}
\end{figure}

We perform a perturbative renormalization at the one-loop level
and find the asymptotic behavior in the $\epsilon$-expansion.
The $\epsilon$-expansion is known as a dimensional regularization for various correlators.
We evaluate the fermion self-energy and the Yukawa interaction 
vertex renormalization as illustrated in Fig.~\ref{Fig4}.
The quantum corrections acquire poles in the $\epsilon \rightarrow 0$ limit,
which drives the one-loop RG running;
more intriguingly, we identify extra $\alpha$-poles
due to the presence of CBS and its curvature effect.

 %--------------------------Fig-------------------------------
%------------------------------------------------------------

\subsection{Fermion self-energy}

We start the one-loop RG analysis for the low-energy theory
by evaluating the fermion self-energy, which is given by
\begin{equation}
\begin{aligned}
\Sigma(p) 
= & -\mu^\epsilon g^2 N_B \int \frac{d^{d+1}q}{(2\pi)^{d+1}} \frac{1}{q_0^2 + q_\parallel^2+\alpha q_\perp^2} \\
& \times \frac{1}{i(p_0+q_0)-v_F\hat{k}_F\cdot ({\bf p}_\perp + {\bf q}_\perp)}.\\
\end{aligned}
\label{eq7}
\end{equation}
where we have abbreviated the momentum variation $\delta q$ to $q$ and
adapt this convention henceforth. 
The extra factor $N_B$ comes from the summation 
over the scattering channels on the critical boson ring.
For a single boson scattering channel near ${\bf q}={\bf q}_B$,
the integral to be evaluated is written as
\begin{equation}
\begin{aligned}
{\cal I}(p)= & \int dq_0dq_\parallel dq_\perp q_\perp^{d-2}\int d\Omega_{d-2} \frac{1}{q_0^2 + q_\parallel^2+\alpha q_\perp^2} \\
& \times \frac{i(p_0+q_0)+(v_F p_\perp^k+ v q_\perp)}{(p_0+q_0)^2+(v_F p_\perp^k+ vq_\perp)^2}, \\
\end{aligned}
\label{eq9}
\end{equation}
where $v=v_F \cos\phi$, and $\phi$ is the $d-2$-dimension polar angle
which illustrated in Fig.~\ref{Fig3}(c). 
$\int d\Omega_{d-2}$ is the $d-2$ dimensional solid angle integral with respect to $\bm{q}_{\perp}$. 
We use the Feynman's parameterization approach
and carry out the integral in the appendix.
The expression of the self-energy correction is
\begin{equation}
\begin{aligned}
\Sigma(p_0,{\bf p}) = \frac{g^2}{4\pi^2v_F} (ip_0) \frac{N_B}{\epsilon\sqrt{\alpha}},
\label{eq:FermionSE}
\end{aligned}
\end{equation}
which corresponds to the fermion wave-function renormalization,
and the Fermi velocity term is not renormalized.
We note that the self-energy takes a different form compared with 
the original fermionic kinetic term $\sim ip_0-v_F p_\perp^k$.
Similar result has been observed in Fermi surface coupled with a single gapless boson\cite{Raghu2014,Gonzalo2014},
which is attributed to the quantum corrections received during renormalization.
The self-energy shows a $\epsilon$-pole in the $\epsilon \rightarrow 0$ as expected.
In addition, there emerges an $\alpha$-divergence which is associated 
with the curvature term of the critical boson ring.
This divergence is enhanced by the number of scattering channels $N_B$ 
which leads to a large factor $N_B/\sqrt{\alpha}\simeq 2\pi q_B^{3/2}/(\Lambda_F\Lambda_B^{1/2})$.
The imaginary part of the fermion self-energy at the Fermi surface
\begin{equation}
{\rm Im}\Sigma(p_0) \sim (g^2N_B/\sqrt{\alpha}) p_0
\label{eq12}
\end{equation}
varies linearly with frequency which is recognized as a marginal NFL behavior\cite{Varma1989}.
In certain perturbative scheme $g^2N_B/\sqrt{\alpha}\sim 1$, 
the quasi-particle decay rate can dominant over the bare energy,
rendering the Landau Fermi liquid picture broken down.
The Yukawa coupling constant is extremely small in the perturbative regime $g_{NFL}\sim (\sqrt{\alpha}/N_B)^{1/2}$, 
thus, the Landau damping energy scale $\omega_{Ld} \sim {\cal O}(g_{NFL}) \sim (\sqrt{\alpha}/N_B)^{1/2} \ll 1$
is suppressed by the large-$N_B$ and curvature effect provided by the CBS.
In a previous study\cite{Raghu2013B}, the Landau damping energy scale receives a 
$N_B^{-1}$ suppression since the Yukawa coupling constant is normalized to be $g\rightarrow g/\sqrt{N_B}$. 
We consider the more physical case with CBS, and the boson polarization function 
is exactly same as the Hertz term written in Ref.(\cite{Raghu2013,Gonzalo2014});
whereas in the NFL regime the $N_B$ factor enters $\omega_{Ld}$ in a different manner.

\subsection{Yukawa interaction vertex}

We evaluate the one-loop renormalization for the Yukawa interaction vertex.
Generically, the Yukawa coupling $\Gamma(k;q)$ is a function of the external Fermi momentum $k$
and the boson momentum transfer $q$. Since the $k$-dependence is continuous 
and we are interested in the low-energy behavior on the Fermi surface,
we set $k_0=0, {\bf k}=k_F{\hat k}_F$ which yields
\begin{equation}
\begin{aligned}
\Gamma(k;q) \simeq & \mu^\epsilon g^3 \int \frac{d^{d+1}p}{(2\pi)^{d+1}} \frac{1}{p_0^2 + p_\parallel^2 + \alpha p_\perp^2}\\
& \times \frac{1}{ip_0 - vp_\perp} \frac{1}{i(p_0+q_0) - (vp_\perp+v_Fq_\perp^k)} \\
\end{aligned}
\label{eq8}
\end{equation}
where the notations are same as Eq.(\ref{eq7}). The correction to the Yukawa vertex only involves a single scattering channel (fixed ${\bf k}_F$ and ${\bf q}_B$) without summation of CBS and there is no $N_B$ here.
In this case, extracting the leading divergence in $\epsilon$
is more involved. We write the detailed calculation in the appendix and present the result here
\begin{equation}
\begin{aligned}
\Gamma(k;q)= & \frac{g^3}{4\pi^2v_F} \frac{iq_0}{iq_0-v_Fq_\perp^k}\frac{1}{\epsilon\sqrt{\alpha}}.\\
\end{aligned}
\end{equation}
We note that the the interaction vertex correction has a $\epsilon$-pole and $\alpha$-divergence,
which are consistent with the result from self-energy. 
One possible explanation for the extra divergence is that
we have adopted two independent UV cutoffs [see around Eq.(\ref{eq6})]
due to the peculiar scattering configuration provided by the infinite critical bosons.
The usual $\epsilon$-pole arises due to the interplay between the small $\epsilon$
and large $k_F\gg \mu\sim \Lambda_F$. The RG scale is related to the cutoff
for the perpendicular components which describes the displacement from the Fermi surface. 
It can be showed that the single $\epsilon$-pole 
is equivalent to a logarithmic divergence of the form $\ln(k_F/\mu)$
once the fermion surface curvature term is included\cite{Gonzalo2014}.
In addition, we have the $\alpha(= \Lambda_B/q_B)$ divergence 
as a ratio between the parallel component and the boson ring radius.
This result reflects the existence of the CBS
and can be attributed to the interplay between the $\epsilon$-expansion and boson curvature term.
In the perturbative regime taken in Eq.(\ref{eq12}), the interaction vertex correction
receives a $1/N_B \sim g^2/\sqrt{\alpha}$ suppression.

The interaction vertex correction shows a non-local dependence on the frequency $q_0$
and the perpendicular component of the momentum fixed by the Fermi vector ${\bf q}_\perp^k$.
The form of the denominator reveals its origin which is related to the linearized fermion Green function,
$iq_0-v_Fq_\perp^k \simeq G_0^{-1}(p) - G_0^{-1}(p+q)$;
physically, this amounts to the particle-hole fluctuation around the Ferm surface.
The interaction vertex and self-energy corrections
satisfy a Ward-identity, 
\begin{equation}
(iq_0-v_Fq_\perp^k)N_B \Gamma(k;q) = g\big[ \Sigma(k+q) - \Sigma(k) \big]
\end{equation}
where we take into account all the scattering channels. The Ward-identity is consistent with the one-loop calculation above.

\subsection{RG flows at one-loop level}

We have evaluated the relevant one-loop order quantum corrections
for the low-energy theory of Fermi surface coupled to the CBS.
The $\epsilon$ poles should be cancelled by the counter-terms defined in Eq.(\ref{eq13}),
which then gives rise to the evolution of RG flow according to Eq.(\ref{eq14}).
The renormalized self-energy amounts to a wave-function counter-term
and the velocity is not renormalized
\begin{equation}
\delta_{\psi} = -\frac{g^2N_B}{4\pi^2v_F}\frac{1}{\epsilon\sqrt{\alpha}}, \ \ \delta_{v}=0 .
\end{equation}
The the anomalous dimension and $\beta$-function for the running velocity are
\begin{equation}
\gamma_{\psi} = \frac{g^2N_B}{8\pi^2v_F \sqrt{\alpha}}, \ \ 
\beta_{v_F} = \frac{g^2N_B}{4\pi^2 \sqrt{\alpha}},
\end{equation}
where we take $\epsilon^{-1} \sim \ln \mu^{-1}$.
The anomalous dimension is a consequence of fermion quasi-particle scattered by gapless bosons;
importantly, this factor receives a significant enhancement in the limit $N_B/\sqrt{\alpha} \rightarrow \infty$,
which is due to the presence of continuous boson minima, namely the CBS in our narrative.
The non-local vertex correction is cancelled by a momentum/frequency dependent counter-term
involving the ratio $q_\perp^k/q_0$,
which leads to
\begin{equation}
\begin{aligned}
\delta_g = -\frac{g^2}{4\pi^2v_F\sqrt{\alpha}}\frac{1}{\epsilon} \frac{1}{1+iv_F(q_\perp^k/q_0)}.
\end{aligned}
\end{equation}
It is quite striking that the counter-term has a dependence on the external momentums.
Such a singular term indicates some soft degree of freedom has been integrated out
and may be attributed to the virtual process of fermions exchanging bosons on the CBS \cite{Gonzalo2014}.

The beta function for Yukawa coupling constant $g$ depends on the running of 
$\delta_\psi$ and $\delta_g$ as dictated in Eq.(\ref{eq14}).
Since $\delta_{\psi}$ receives an additional large-$N_B$ factor compared to $\delta_g$,
the dominant contribution to $\beta_g$ comes from the fermion wave-function renormalization,
namely we have
\begin{equation}
\beta_g = -\frac{\epsilon}{2}g + \frac{g^3N_B}{4\pi^2v_F\sqrt{\alpha}}.
\label{eq15}
\end{equation}
This leads to a large-$N_B$ driven NFL fixed solution $\beta_{g^\ast}=0$ at
\begin{equation}
g^\ast = \sqrt{\frac{4\pi^2v_F\sqrt{\alpha}\epsilon}{N_B}}.
\end{equation}
This is the main result of this section. We give a few remarks on the role of
the large number of bosons on CBS.
It's noted that the fixed value is suppressed by a factor $N_B^{-1/2}$;
more importantly, the singular momentum dependence of $\delta_g$ 
usually causes problem for the RG equation, which is avoided here due to
the large-$N_B$ flavor of critical bosons.
To get more insights of this problem, let's set $N_B=1$ by hand at this step, which would yield 
\begin{equation}
\beta_g^{N_B=1} = -\frac{\epsilon}{2}g + \frac{g^3}{4\pi^2v_F\sqrt{\alpha}} \frac{1}{1-i(q_0/q_\perp^k)/v_F}.
\end{equation}
The result is quite different from Eq.(\ref{eq15}) since the flows of $\delta_g$ and $\delta_\psi$ 
both contribute to the beta function.
As mentioned previously, the perpendicular component $q_\perp^k$
dictates the displacement from the nearest Fermi surface point, 
therefore, is bounded by $\Lambda_F\sim \mu$.
We can parameterize the ratio by the RG scale by setting $q_0=x\mu, \ q_\perp^k=\mu$\cite{Gonzalo2014}
and the ratio $x$ is left to be a free parameter, 
we rely on tuning this parameter to reach certain fixed points.
In the $x\rightarrow 0$ limit,
we obtain a flow equation 
\begin{equation}
\beta_{g}^{N_B=1} = -\frac{\epsilon}{2}g + \frac{{g}^3}{4\pi^2v_F\sqrt{\alpha}} =0
\end{equation}
which gives rise to a NFL fixed point at $g^\ast={\cal O}(\sqrt{\epsilon\sqrt{\alpha}})$.
On the other hand, the RG flow is slowly running $\beta_g = -\frac{\epsilon}{2}g$
in the limit $x\rightarrow \infty$, and higher-loop corrections have to be taken into
consideration, which may leads to a perturbative fixed point.
The physical property depends on the limit of $x$ we choose respecting the non-local contribution generated under renormalization.

\section{Discussion}
\label{sec4}
We study a fermion-boson coupled system 
with both Fermi surface and critical boson surface in $d=3-\epsilon$ spatial dimensions.
An effective theory is developed for the quantum critical metal 
above the Landau damping energy scale $\omega> \omega_{Ld}$.
In such a regime, the theoretical approach differs from the well-known Hertz-Millis method.
We carry out an $\epsilon$-expansion perturbatively up to one-loop RG,
which amounts to evaluating the flows of the Yukawa coupling constant 
and the fermion anomalous dimension.
We discover a NFL fixed point that exists particularly in the large-$N_B$ limit
and further compare this result to the usual $N_B=1$ case.

The infinite many critical bosons on a sphere
introduce a large number of inelastic scattering channels for fermions.
These channels play a similar role as the flavor degree of freedom in the matrix large-$N_B$ limit,
yet, affect the RG flow equations in a distinct way.
In both cases, the bosons outnumber the fermions,
rendering the singular Landau damping effect largely suppressed for $\omega>\omega_{Ld}$.
In the opposite limit $\omega<\omega_{Ld}$, we tend to believe that 
the Hertz-Millis approach still works\cite{Zhang2021} based on the theoretical ground.
Various theoretical results are obtained from different models and RG schemes,
which are left to be examined by experiments;
for instance the imaginary self-energy is related to the resistivity in the transport experiment.
To fully determine the role of Landau damping effect,
one can include the boson polarization generated at one-loop (which reduces to the Hertz term in certain dynamical limit)
to the original theory and re-analyze the fermion self-energy\cite{Gonzalo2014}.
This will change the dynamical exponent $z$ and modifies the low-energy physics a lot.

In the current theory, the pure four fermion interaction is irrelevant at $d=3$ and omitted.
This will not be the case in the lower dimensions.
A further comprehensive theory should take into account the four fermion interaction. 
In the fermionic renormalization group theory (Shankar's RG\cite{Shankar1994}), 
four fermion interaction can be separated into different channels, 
where forward scattering and BCS pairing described by Landau parameters are most relevant.
The possible Fermi surface instabilities could be enhanced by the gapless boson mode.
In the series works\cite{Gonzalo2014,Gonzalo2015}, 
a tree-level running of four-fermion interaction was found, 
which is necessary for the renormalizability of the theory.
The BCS superconducting pairing could be enhanced near the quantum critical point,
which is expected to be more prominent for our case with infinite many critical bosons.
Moreover, the existence of CBS calls for more insight into boson quartic self-interaction, 
e.g., separating of channels like the four fermion interaction.
In total, the theory has several energy scales, including 
the Landau-damping scale $\omega_{Ld}$, the BSC pairing scale and the NFL scale.
A complete phase diagram could be obtained if all these are considered, which we leave for the further works.

\section*{Acknowledgments} 
We thank Gonzalo Torroba and Huajia Wang for discussion. 
X.T.Z. acknowledges Prof. Gang Chen at the University of Hong Kong 
for collaboration on previous related projects.
X.T.Z. is grateful for the hospitality of Shenghan Jiang 
at Kavli Institute of Theoretical Sciences (Beijing,China)
during the visiting period of time this work was motivated.
X.T.Z. is supported by the National Science Foundation of 
China with Grant No.~92065203, the Ministry of Science and 
Technology of China with Grants No.~2021YFA1400300 
and by the Research Grants Council of Hong Kong 
with General Research Fund Grant No. 17306520.
Z.P. is supported by National Natural Science Foundation of China (No. 12147104).

\onecolumngrid
\appendix
\section{Dimensional regularization in asymptotic $\epsilon$-expansion
and $\alpha \rightarrow 0$}

In the appendix, we give a step-by-step derivation for the epxression
of self-energy and interaction vertex correction presented in Sec.~\ref{sec3}.
To this end, we find the following relation to be useful in evaluating the integrals encountered later.
The primary relation is
\begin{equation}
\begin{aligned}
I(n,d) \equiv &\int dq_0 dq_\parallel \int_0^{\infty}dq_\perp q_\perp^{d-2}  \frac{1}{\big[Aq_0^2 + B q_\parallel^2 + Cq_\perp^2 + \Delta\big]^n}
= \frac{\pi}{2} \frac{\Gamma\big[\frac{2n-d-1}{2}\big] \Gamma\big[\frac{d-1}{2}\big]}{\Gamma(n)} \frac{1}{\sqrt{ABC^{d-1}\Delta^{2n-d-1}}} \\
\end{aligned}
\label{A_eq1}
\end{equation}

We can take a derivative with respect to coefficients $\frac{dI(n,d)}{dA}$, which yield another useful relation
\begin{equation}
\begin{aligned}
& \int dq_0 dq_\parallel \int_0^{+\infty} dq_\perp q_\perp^{d-2} \frac{q_0^2}{\big[Aq_0^2 + B q_\parallel^2 + Cq_\perp^2 + \Delta\big]^n}
= \frac{\pi}{4}\frac{\Gamma[\frac{2n-d-3}{2}]\Gamma[\frac{d-1}{2}]}{\Gamma(n)} \frac{1}{\sqrt{A^3BC^{d-1}\Delta^{2n-d-3}}} \\
\end{aligned}
\label{A_eq2}
\end{equation}

Similarly, we also have
\begin{equation}
\begin{aligned}
&\int dq_0 dq_\parallel \int_0^{\infty}dq_\perp q_\perp^{d-2}  \frac{q_{\perp}^2}{\big[Aq_0^2 + B q_\parallel^2 + Cq_\perp^2 + \Delta\big]^{n}}
=\frac{\pi}{2} \frac{d-1}{2} \frac{\Gamma\big[\frac{2n-d-3}{2}\big] \Gamma\big[\frac{d-1}{2}\big]}{\Gamma(n)} \frac{1}{\sqrt{ABC^{d+1}\Delta^{2n-d-3}}}
\end{aligned}
\label{A_eq3}
\end{equation}

\subsection{Self-energy}

Armed with the relations, we calculate the integral in Eq.(\ref{eq9})
using the Feynman's parameterization approach.
The denominator is transformed as, 
\begin{equation}
\begin{aligned}
& \frac{1}{q_0^2 + q_\parallel^2+\alpha q_\perp^2} \frac{1}{(p_0+q_0)^2+(v_F p_\perp^k+ vq_\perp)^2}
= \frac{\Gamma(2)}{\Gamma^2(1)}\int_0^1 dx \frac{1}{\Big\{(1-x)(q_0^2 + q_\parallel^2) + x\big[(p_0+q_0)^2+(v_F p_\perp^k+ vq_\perp)^2\big] \Big\}^2} \\
= & \int_0^1 dx \frac{1}{\Big\{ q_0^2 + 2xp_0 q_0 + (1-x)q_\parallel^2 +[\alpha (1-x)+ xv^2]q_\perp^2 + 2 xv_F^2 p_\perp^k\cos\phi q_\perp +x\big[p_0^2+v_F^2(p_\perp^k)^2\big] \Big\}^2} 
\end{aligned}
\end{equation}

We make a change of variable
\begin{equation}
\begin{aligned}
q_0^\prime = & q_0 + xp_0\\
q_\perp^\prime = & q_\perp + \frac{xv_F^2\cos\phi  }{\alpha (1-x)+ xv^2} p_\perp^k\\
\end{aligned}
\end{equation}
then, the integral in Eq.(\ref{eq9}) reads
\begin{equation}
\begin{aligned}
{\cal I}(q)
= &  \int dq_0 dq_\parallel d^{d-1}q_\perp \int_0^1 dx \frac{i\big[ (1-x)p_0 + q_0\big]+ (v_F p_\perp^k)\frac{\alpha(1-x)}{\alpha (1-x)+ xv^2}+ vq_\perp}{\big[Aq_0^2 + B q_\parallel^2 + Cq_\perp^2 + \Delta\big]^2} ,\\
= & (ip_0) \int_0^1 dx(1-x) \int d^{d-1}q_\perp dq_0 dq_\parallel  \frac{1}{\big[Aq_0^2 + B q_\parallel^2 + Cq_\perp^2 + \Delta\big]^2} \\
& + v_F \int_0^1 dx \int d^{d-1}q_\perp dq_0 dq_\parallel  \frac{\cos\phi q_\perp}{\big[Aq_0^2 + B q_\parallel^2 + Cq_\perp^2 + \Delta\big]^2}  \\
& + \alpha (v_Fp_\perp^k) \int_0^1 dx(1-x) \int d^{d-1}q_\perp dq_0 dq_\parallel \frac{1}{\alpha (1-x)+ xv^2} \frac{1}{\big[Aq_0^2 + B q_\parallel^2 + Cq_\perp^2 + \Delta\big]^2} \ ,\\
\equiv & {\cal I}_1 + {\cal I}_2 + {\cal I}_3. \\
\end{aligned}
\label{A_I}
\end{equation}
We have dropped the odd term. The coefficients are given by
\begin{equation}
\begin{aligned}
& A =1, \ \ B= 1-x, \ \ C= \alpha(1-x) +xv^2, \\
& \Delta = x(1-x) \Big\{ p_0^2 + \alpha \frac{1}{\alpha(1-x) +xv^2} (v_F p_\perp^k)^2\Big\}
\simeq  x(1-x) p_0^2.\\
\end{aligned}
\end{equation}
where we take $\alpha\rightarrow 0$ limit in the last equation.

We calculate the integrals in Eq.(\ref{A_I}) separately.
Firstly, the ${\cal I}_1$-integral is evaluated using the relation in Eq.(\ref{A_eq1}) with $n=2, d=3-\epsilon$, which yields
\begin{equation}
\begin{aligned}
{\cal I}_1(p)
= & (ip_0)\epsilon^{-1}\pi \int d^{d-2}\Omega \int_0^1 dx (1-x)
\frac{1}{(1-x)^{1/2}} \frac{1}{\big[ \alpha(1-x) +xv^2 \big]^{1-\epsilon/2}} \frac{1}{\Delta^{\epsilon/2}}  \\
\end{aligned}
\label{A_I1}
\end{equation}

Recall that $v=v_F \cos\phi$ and
the $\phi$-integral is calculated as
\begin{equation}
\begin{aligned}
& \int_0^{2\pi} d\phi \frac{1}{\big[ \alpha(1-x) +xv^2 \big]^{1-\epsilon/2}}
= \frac{1}{v_F^2}\frac{1}{x^{1-\epsilon/2}} \Big\{ \frac{2\pi}{\sqrt{a}} \frac{1}{\sqrt{1+a}}+ o(\epsilon) \Big\}
\simeq \frac{1}{v_F^2}\frac{1}{x^{1-\epsilon/2}} \frac{2\pi}{\sqrt{a}} \\
\end{aligned}
\end{equation}
where the parameter is defined as $a= \alpha v_F^{-2} (1-x)/x$,
and we have adopted the $\alpha \rightarrow 0$ limit.
Finally, the $x$-integral is carried out
\begin{equation}
\begin{aligned}
{\cal I}_1(p)
= & (ip_0)\epsilon^{-1} \frac{2\pi^2}{\sqrt{\alpha} v_F}
\int_0^1 dx \frac{1}{x^{1/2-\epsilon/2}} \frac{1}{[x(1-x)]^{\epsilon/2}} 
\simeq \frac{(2\pi)^2}{v_F}(ip_0)\frac{1}{\epsilon\sqrt{\alpha}}   \\
\end{aligned}
\end{equation}

We note that the proper sequence for the integrals is crucial;
if the $x$-integral is carried out first, the subsequent $\phi$-integral gives a divergent result.\\

Secondly, the ${\cal I}_2$-integral is evaluated using the relation in Eq.(\ref{A_eq2}) with $n=2, d=3-\epsilon$, 
which yields a vanishing result
\begin{equation}
\begin{aligned}
{\cal I}_2(p)
= & v_F \int_0^1 dx \int d^{d-2}\Omega \cos\phi \int dq_0 dq_\parallel\int_0^{+\infty} dq_\perp  
\frac{q_\perp^{d-1}}{\big[Aq_0^2 + B q_\parallel^2 + Cq_\perp^2 + \Delta\big]^2} \\
= & -v_F \frac{\pi^2}{2} \int_0^{2\pi} d\phi \cos\phi  \int_0^1 dx \frac{1}{(1-x)^{1/2}} 
\frac{1}{\big[\alpha(1-x)+xv^2\big]^{(3-\epsilon)/2}} \frac{1}{\big[ x(1-x)\big]^{-(1-\epsilon)/2}} \\
= & 0  \\
\end{aligned}
\end{equation}

Finally, the ${\cal I}_3$-integral also has a vanishing result
which can be verified by setting $\alpha=0$.
Note that this vanishing result suggests that 
the self-energy has no external $p_\perp^k$ dependence.

\subsection{Yukawa interaction vertex}

Next, we evaluate the Yukawa interaction vertex by fixing the 
external Fermi momentum as $k=(k_0=0,{\bf k}=k_F \hat{k}_F)$,
\begin{equation}
\begin{aligned}
\Gamma(k;q)
= &  \mu^\epsilon g^3 \int \frac{d^{d+1}p}{(2\pi)^{d+1}} \frac{1}{p_0^2 + p_\parallel^2 + \alpha p_\perp^2} 
 \frac{ip_0 + v_Fp_\perp^k}{p_0^2 +v^2 p_\perp^2} 
\frac{i(p_0+q_0) + (vp_\perp+v_Fq_\perp^k)}{(p_0+q_0)^2+ (vp_\perp+v_Fq_\perp^k)^2} \\
\end{aligned}
\label{A_eq3}
\end{equation}
The denominators are combined using the Feynman's parameterization
\begin{equation}
\begin{aligned}
\sim &(1-x-y)(p_0^2 + p_\parallel^2 + \alpha p_\perp^2) + y(p_0^2 +v^2p_\perp^2) 
+ x \big[(p_0+q_0)^2+ (vp_\perp+ v_Fq_\perp^k)^2\big] \\
= & (p_0 + \delta_0)^2 + (1-x-y)p_\parallel^2 
 + \big[\alpha(1-x-y)+ (x+y)v^2\big](p_\perp+\delta_\perp)^2 + \Delta(x,y),\\
\end{aligned}
\end{equation}
where we define
\begin{equation}
\begin{aligned}
& \delta_0 = xq_0, \ \ \delta_\perp = \frac{xv_F v q_\perp^k}{\alpha(1-x-y)+ (x+y)v^2},\\
& \Delta(x,y) =x(1-x)q_0^2 + (v_F{q_\perp^k})^2 \frac{x(1-x ) v^2}{\alpha(1-x-y)+ (x+y)v^2}. \\
\end{aligned}
\end{equation}

The integral involved in Eq.(\ref{A_eq3}) is defined and calculated as
\begin{equation}
\begin{aligned}
{\cal I}(q) \equiv & 2\int_{0}^{1} dx \int_{0}^{1-x}dy \int d^{d+1}p
 \frac{\big[ i(p_0+q_0-\delta_0) + (vp_\perp+v_Fq_\perp^k-v\delta_\perp)\big]\big[i(p_0-\delta_0) + (vp_\perp-v\delta_\perp)\big]}{\big\{p_0^2 + (1-x-y)p_\parallel^2 + \big[\alpha(1-x-y)+ (x+y)v^2\big]p_\perp^2 + \Delta(x,y) \big\}^3} \\
= & 2\int_{0}^{1} dx \int_{0}^{1-x} dy \int_0^{2\pi}d\phi \int dp_0dp_\parallel d^{d-1}p_\perp 
\frac{(-p_0^2+v^2p_\perp^2) + A(x,y)}{\big\{p_0^2 + (1-x-y)p_\parallel^2 + \big[\alpha(1-x-y)+ (x+y)v^2\big]p_\perp^2 + \Delta(x,y) \big\}^3}\\
\equiv &{\cal I}_1(q) + {\cal I}_2(q).  \\
\end{aligned}
\end{equation}
The integrals have two terms which are calculated separately,
and the function is given by
\begin{equation}
\begin{aligned}
A(x,y) = & -\big[i(q_0-\delta_0) +(v_Fq_\perp^k-v\delta_\perp)\big](i\delta_0+v\delta_\perp) \\
\end{aligned}
\end{equation}

The ${\cal I}_1$ integral is transformed and further divided into two pieces,
\begin{equation}
\begin{aligned}
{\cal I}_1(q) = &2\int_{0}^{1} dx \int_{0}^{1-x} dy  \int_0^{2\pi}d\phi \frac{\pi}{4}\frac{\Gamma(\epsilon/2)\Gamma[(2-\epsilon)/2]}{\Gamma(3)}\frac{\big[\alpha(1-x-y)+ (x+y)v^2\big]^{\epsilon/2}}{(1-x-y)^{1/2} \Delta^{\epsilon/2}}  \\
& \times \Big\{-\frac{1}{\big[\alpha(1-x-y)+ (x+y)v^2\big]} + \frac{(d-1)v^2}{\big[\alpha(1-x-y)+ (x+y)v^2\big]^2}\Big\},\\
= &  \frac{1}{\epsilon}\frac{\pi}{2} \int_{0}^{1} dx \int_{0}^{1-x} dy\int_0^{2\pi}d\phi\frac{\big[\alpha(1-x-y)+ (x+y)v^2\big]^{\epsilon/2}}{\Delta^{\epsilon/2}}\Big\{ \frac{-(\alpha-v^2)(1-x-y)^{1/2} }{\big[\alpha(1-x-y)+ (x+y)v^2\big]^2} \\
& + \frac{(d-2)}{(1-x-y)^{1/2}\big[\alpha(1-x-y)+ (x+y)v^2\big]^2} \Big\}  \\
\equiv & {\cal I}_1^a + {\cal I}_1^b,
\end{aligned}
\end{equation}
where the parameter is given by
\begin{equation}
a= \alpha v_F^{-2} \frac{1-x-y}{x+y} 
\end{equation}

Part $a$ is calculated as
\begin{equation}
\begin{aligned}
{\cal I}^a_1(q)
\simeq &  \frac{1}{\epsilon}\frac{\pi}{2} \int_{0}^{1} dx \int_{0}^{1-x} dy (1-x-y)^{1/2} 
 \int_0^{2\pi}d\phi\frac{-(\alpha-v^2)}{\big[\alpha(1-x-y)+ (x+y)v^2\big]^2}  \\
= & \frac{1}{\epsilon}\frac{\pi}{2}  \int_{0}^{1} dx \int_{0}^{1-x} dy\frac{(1-x-y)^{1/2}}{(x+y)^2v_F^4} 
\frac{\pi}{a^{1/2}(1+a)^{3/2}}\Big\{ -\alpha\frac{1+2a}{a} + v_F^2 \Big\}   \\
\simeq & \frac{1}{\epsilon}\frac{\pi}{2} \int_{0}^{1} dx \int_{0}^{1-x} dy \frac{(1-x-y)^{1/2}}{(x+y)^2v_F^4}  \frac{\pi v_F^2}{a^{1/2}}\Big\{ 2- \frac{1}{1-x-y} \Big\}  \\
= & \frac{1}{\epsilon \sqrt{\alpha}} \frac{2\pi^2}{v_F}- \frac{1}{\epsilon \sqrt{\alpha}} \frac{\pi^2}{2v_F}  \int_{0}^{1} dx \int_{0}^{1-x} dy\frac{1}{(x+y)^{3/2}(1-x-y)}. \\
\end{aligned}
\label{A_eq4}
\end{equation}
where the 2nd term in the last equality is divergent upon integrating over $x,y$.
Fortunately, this term is cancelled by the part $b$ as we show below,
\begin{equation}
\begin{aligned}
{\cal I}_1^{b}(q)
\simeq&\frac{\pi}{2\epsilon} \int_{0}^{1} dx \int_{0}^{1-x} dy  \int_0^{2\pi}d\phi \frac{1}{(1-x-y)^{1/2} }
\frac{v^2}{\big[\alpha(1-x-y)+ (x+y)v^2\big]^{2}} \\
=&\frac{1}{\epsilon} \frac{\pi}{2}  v_F^{-2} \int_{0}^{1} dx \int_{0}^{1-x} dy  \int_0^{2\pi}d\phi
\frac{1}{(x+y)^2 (1-x-y)^{1/2} } \frac{\pi}{ \sqrt{a} (1+a)^{3/2}} \\
\simeq &\frac{1}{\epsilon \sqrt{\alpha}} \frac{\pi^2}{2v_F}  \int_{0}^{1} dx \int_{0}^{1-x} dy  \frac{1}{(x+y)^{3/2} (1-x-y) }  \\
\end{aligned}
\label{A_eq5}
\end{equation}

Combining the two pieces from Eq.(\ref{A_eq4}) and Eq.(\ref{A_eq5}),
we obtain 
\begin{equation}
\begin{aligned}
{\cal I}_1(q)=\frac{1}{\epsilon \sqrt{\alpha}} \frac{2\pi^2}{v_F}
\end{aligned}
\label{A_eq6}
\end{equation}

The ${\cal I}_2$-integral is quite complicated,
\begin{equation}
\begin{aligned}
{\cal I}_2(q)
\simeq & \int_{0}^{1} dx dy \int_0^{2\pi}d\phi A(x,y)\frac{1}{(1-x-y)^{1/2}}
\frac{1}{\big[\alpha(1-x-y)+ (x+y)v^2\big]^{1-\epsilon/2}}\frac{1}{\Delta^{1+\epsilon/2}}\
\end{aligned}
\end{equation}
and the purpose here is to extract the leading $\epsilon$ divergence.
We find it helpful to take the $y\rightarrow 1-x$ limit,
where the functions happen to take simply forms
\begin{equation}
\begin{aligned}
& \Delta(x,1-x) = x(1-x)(q_0^2+v_F^2{q_\perp^k}^2) \\
& A(x,1-x) =-x(1-x) (iq_0+v_Fq_\perp^k)^2 \\
\end{aligned}
\end{equation}
and we still integrate over $x,y$ variables for the remaining parts of the integrand 
\begin{equation}
\begin{aligned}
{\cal I}_2(q)
\simeq & \frac{\pi}{2}\int_{0}^{1} dx \int_0^{1-x}dy \int_0^{2\pi}d\phi \frac{1}{(1-x-y)^{1/2}}
\frac{1}{\big[\alpha(1-x-y)+ (x+y)v^2\big]^{1-\epsilon/2}} \frac{iq_0+v_Fq_\perp^k}{iq_0-v_Fq_\perp^k} \\
= & \frac{iq_0+v_Fq_\perp^k}{iq_0-v_Fq_\perp^k}\frac{\pi}{2v_F^2}\int_{0}^{1} dx \int_0^{1-x}dy 
\frac{1}{(1-x-y)^{1/2}(x+y)^{1/2}}\frac{1}{a^{-\epsilon/2}} \frac{2\pi}{\sqrt{a}\sqrt{1+a}}\\
\simeq &\frac{iq_0+v_Fq_\perp^k}{iq_0-v_Fq_\perp^k} \frac{1}{\sqrt{\alpha}} \frac{\pi^2}{v_F}\int_{0}^{1} dx \int_0^{1-x}dy
\frac{1}{(1-x-y)^{1-\epsilon/2}(x+y)^{\epsilon/2}} \\
\simeq & \frac{1}{\epsilon\sqrt{\alpha}} \frac{2\pi^2}{v_F}  \frac{iq_0+v_Fq_\perp^k}{iq_0-v_Fq_\perp^k}\\
\end{aligned}
\label{A_eq7}
\end{equation}

Collecting terms from Eq.(\ref{A_eq6}) and Eq.(\ref{A_eq7}),
we end up with the result for Yukawa interaction vertex correction
\begin{equation}
\begin{aligned}
\Gamma(k;q) 
= &  \frac{4\pi^2 g^3}{v_F} \frac{iq_0}{iq_0-v_Fq_\perp^k} \frac{1}{\epsilon\sqrt{\alpha}}\\
\end{aligned}
\end{equation}

\twocolumngrid
\bibliography{Ref.bib}

\end{document}